\documentclass[useAMS]{mn2e}
\def\i{f_{24\mu{\rm m}}}
\def\r{f_{20{\rm cm}}}

\usepackage{graphicx}

\title{Extending the infrared radio correlation}
\author[B.J.Boyle et al.]
	{B.J.Boyle$^{1}$, T.J. Cornwell$^{1}$, E.Middelberg$^{1}$,     
	R.P.Norris$^{1}$, P.N.Appleton,$^{2}$, Ian Smail$^{3}$\\
$^{1}$ Australia Telescope National Facility, PO Box 76, Epping, NSW 1710, Australia\\
$^{2}$ {\it Spitzer} Science Center, California Institute of Technology, 1200 E.California Blvd., Pasadena, CA 91125, USA\\
$^{3}$ Institute for Computational Cosmology, University of Durham, South Road, Durham DH1 3LE, UK\\
}

\begin{document}

\maketitle

\pagerange{\pageref{firstpage}--\pageref{lastpage}} \pubyear{2004}

\label{firstpage}

\begin{abstract}
Co-addition of deep (rms $\sim 30 \mu$Jy) 20cm data obtained with the
Australia Telescope Compact Array at the location of {\it Spitzer} Wide
field survey (SWIRE) sources has yielded statistics of radio source
counterparts to faint 24$\mu$m sources in stacked images with rms
$<1\mu$Jy.  We confirm that the infrared-radio correlation extends to
$\i=100\mu$Jy but with a significantly lower coefficient,
$\r=0.039\i$ ($q_{24} =  \log(\i/\r) = 1.39\pm0.02$) than hitherto reported.  We postulate that this may be
due to a change in the mean $q_{24}$ value ratio for objects with $\i<1$mJy.
\end{abstract}



\section{Introduction}

The tight correlation between the global far-infrared and radio
emission from galaxies has been established for over three decades.
Early ground-based studies (van der Kruit 1973; Condon et al.\ 1982)
and observations with the infrared Astronomical Satellite ({\it IRAS},
Dickey \& Salpeter 1984; de Jong et al.\ 1985) conclusively
established a correlation over a broad range of Hubble types and
luminosities at low redshifts ($z<0.1$).  The origin of this
correlation is thought to lie in the link between massive stars, which
generate far-infrared emission by reheating dust, and supernovae,
which accelerate cosmic rays that then generate radio synchrotron
radiation (Harwit \& Pacini 1975; Condon, 1992; Xu, Lisenfield \&
Volk 1994).  However, alternative mechanisms have also been proposed,
including secondary electron production (radio synchrotron) from
molecular clouds (far infrared).

More recently, by extrapolating observations made with the Infrared
Space Observatory ({\it ISO}) in the mid-infrared (Cohen et al.\ 2000,
Garrett 2002) and in the sub-millimetre with the SCUBA instrument,
(Carilli \& Yun 1999, Ivison et al.\ 2002), it has been argued that the
far-infrared/radio correlation extends to $z>1$.  If confirmed, this
suggests that the galaxies at high redshift may share many of the same
properties as their low redshift counterparts; whatever the origin of
this relation.

With the greatly enhanced mid-infrared sensitivity provided by the
{\it Spitzer} Satellite, it is now possible to investigate directly the
far-infrared/radio correlation at redshifts of order unity.  An
initial attempt has already been made by Appleton et al.\ (2004),
using the {\it Spitzer} First Look Survey (FLS) and the Very Large Array
(VLA). The flux limits for the FLS and VLA samples were $\i=0.3$mJy
and $\r=115$mJy respectively.  Although limited by the availability of
spectroscopic redshifts, Appleton et al. were able to demonstrate a
good correlation between the 24$\mu$m and 20cm flux for objects with a
median $z = 0.3$ extending out to $z\sim 1$.  Their analysis was
limited by the relative brightness of the radio flux limit, which was
a factor of three brighter than that corresponding to the 24$\mu$m
limit, based on the observed far-infrared/radio flux ratio.

In this analysis, we extend the flux limits to which the
far-infrared/radio correlation may be studied by a further factor of
three at 24$\mu$m and by a factor of 20 at 20\,cm, using the {\it Spitzer}
Wide Field Survey (SWIRE) and the technique of stacking (see e.g. Wals
et al.\ 2005) applied to deep Australia Telescope Compact Array (ATCA)
observations.

\begin{figure*}
\centering
\includegraphics[scale=0.6]{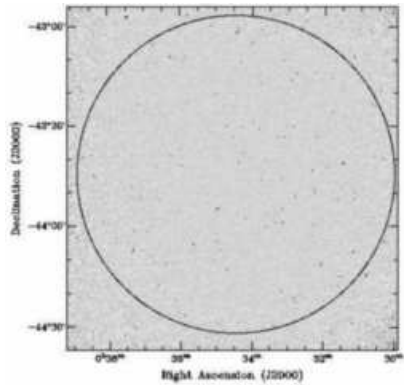}
\caption{ATCA 20cm images of the CDFS field (left) and the ELAIS field
(right).  The areas used in these analyses is denoted by the circles.}
\label{cdfs}
\end{figure*}

\section{Data}

\subsection{ATCA observations}

Two regions around the Chandra Deep Field South (CDFS,
RA(2000)=$3^{\rm h}31^{\rm m}$ Dec(2000)=$-28^{\circ}06^{\rm m}$) and the
European Large Area {\it ISO} Survey S1 (ELAIS, RA(2000)=0h34,
DEC(2000)=$-43^{\circ}45^{\rm m}$, have been observed with a series of
mosaiced pointings at 20cm with the ATCA as part of the Australia
Telescope Large Area Survey (ATLAS, Norris et al.\ 2006; Middelberg et
al.\ 2007, in prep). The goal of the ATLAS program is to provide large
area (6 deg$^2$), deep surveys of radio sources in fields which are
also the subject of intensive study at other wavebands. In particular,
the areas covered are chosen to be matched to those covered by the
SWIRE survey (Lonsdale et al.\ 2003) as well as enabling optical
spectroscopy with instruments such as the Anglo Australian Telescope
2-degree field. A total of 360 hours integration has been obtained on
the CDFS, and of 234 hours in the ELAIS. In each field, 28 and 20
pointings (each with a mean integration of 13 hours) have been
mosaiced together to provide an area of approximately 1.6 deg $\times$
2.2\ deg in the CDFS, and of 1.6\,deg $\times$ 1.6\,deg in the ELAIS.

The data were calibrated and imaged using Miriad (Sault et al.\ 1995) following the procedures in the Miriad cookbook for high dynamic range observations. In particular, multi-frequency clean had to be used to account for the high fractional bandwidth of 15 per cent.  The mean rms noise in the images is 30$\mu$Jy, and is limited not by integration time but by sidelobes arising from strong sources in (CDFS) or just outside (ELAIS) the imaged areas (shown in Fig~\ref{cdfs}). The CDFS image also contains data from an earlier observation by Koekemoer et al. (in preparation), who have covered a region of 1 deg diameter centred on RA(2000)=3$^{\rm h}$33, Dec(2000)=$-27^{\circ}$ with a mosaic of seven pointings, and reach an rms noise level of $\sim 15\mu$Jy. In order to avoid regions around the edge of the mosaic with increased noise levels, we restricted the analysis to a 48-arcmin radius centred on the ATLAS field centres, denoted with circles in Fig~\ref{cdfs}.  As a further check of the calibration, we confirmed that there was no significant offset between the published NVSS fluxes and the flux estimates obtained from the ATCA data for bright sources ($\r>2$\,mJy) in the CDFS field (see also Norris et al.\ 2006)

\subsection{{\it Spitzer} observations}

The {\it Spitzer} Space Telescope was launched in August 2003. It is
equipped with several instruments, and here we use data taken with the
Multiband Imaging Photometer for {\it Spitzer} (MIPS) in its 24$\mu$m band.
Before launch, proposals were invited for Legacy Science Programs,
which were large coherent programs whose data would be of lasting
importance to the broad astronomical community.  One of the six Legacy
projects chosen was the SWIRE ({\it Spitzer} Wide Extragalactic) Survey
(Lonsdale et al.\ 2003), which has observed a region of 6 square
degrees surrounding the CDFS, and of 5 square degrees coincident with
the ELAIS field. The analysis of those data are described by Surace et al.\
(2006). Most of the SWIRE data are now in the public domain (SWIRE
data Release 3\footnote{http://swire.ipac.caltech.edu/swire/astronomers/data\_access.html}). Here
we use only the 24$\mu$m MIPS data. The fluxes used here are, in the case of unresolved or noisy sources, aperture-corrected fluxes, as described by Surace et al.\ (2006). In the case of extended sources, Kron fluxes are used (Kron 1980).

\begin{figure}
\centering
\includegraphics[scale=0.5,ext=ps]{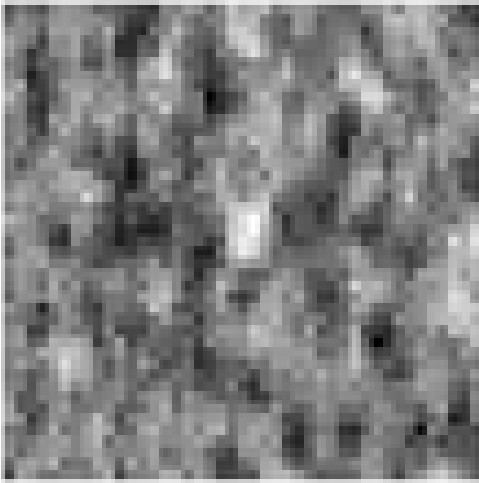}
\includegraphics[scale=0.5,ext=ps]{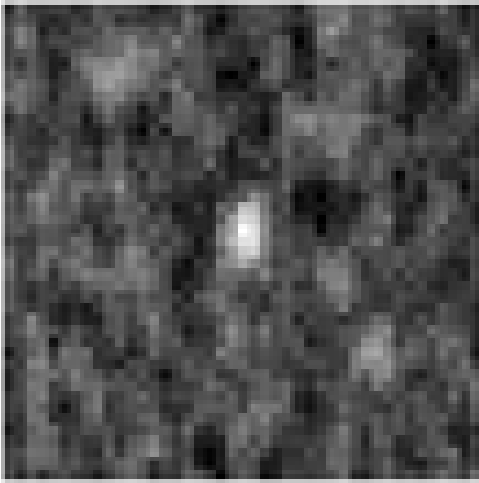}
\includegraphics[scale=0.5,ext=ps]{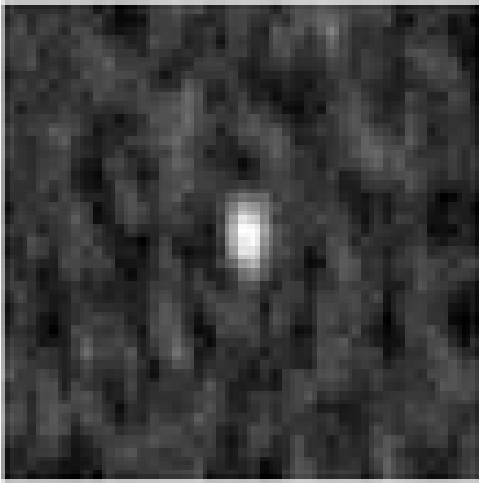}
\includegraphics[scale=0.5,ext=ps]{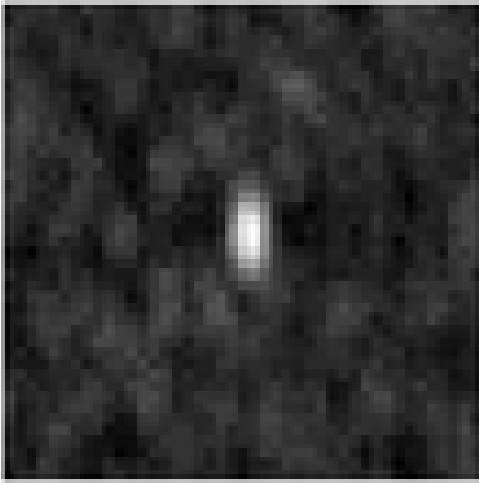}
\includegraphics[scale=0.5,ext=ps]{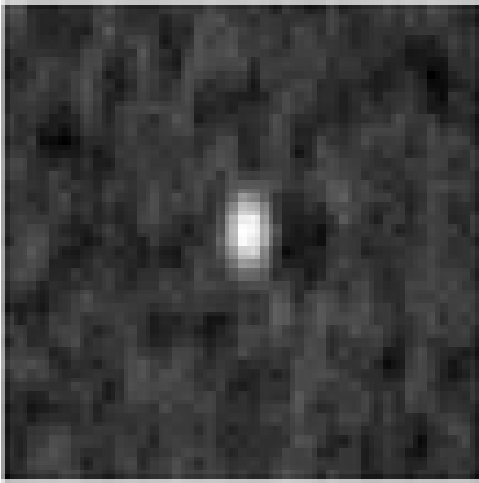}
\includegraphics[scale=0.5,ext=ps]{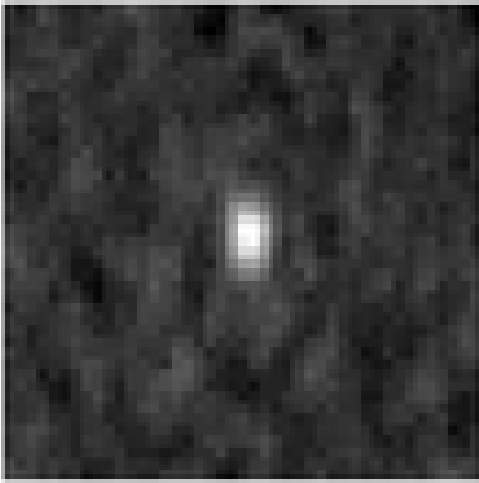}
\includegraphics[scale=0.5,ext=ps]{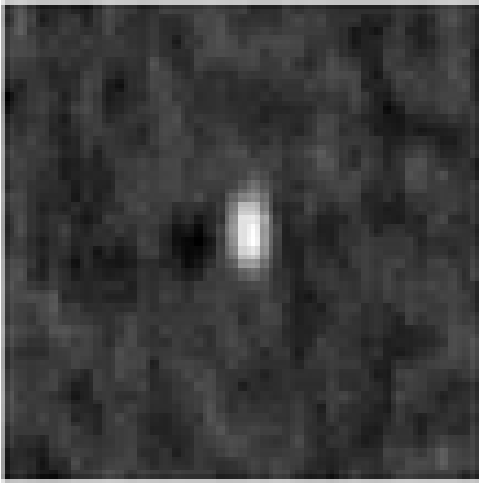}
\includegraphics[scale=0.5,ext=ps]{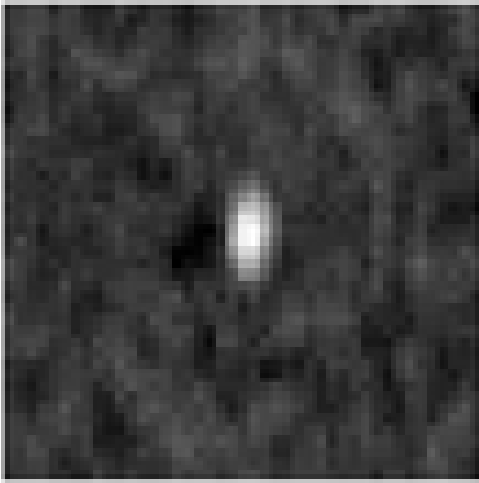}
\includegraphics[scale=0.5,ext=ps]{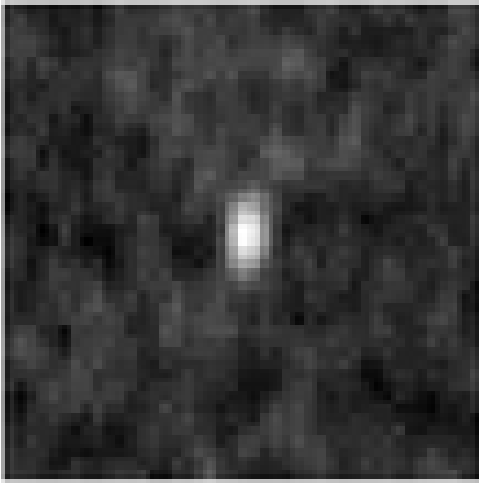}
\includegraphics[scale=0.5,ext=ps]{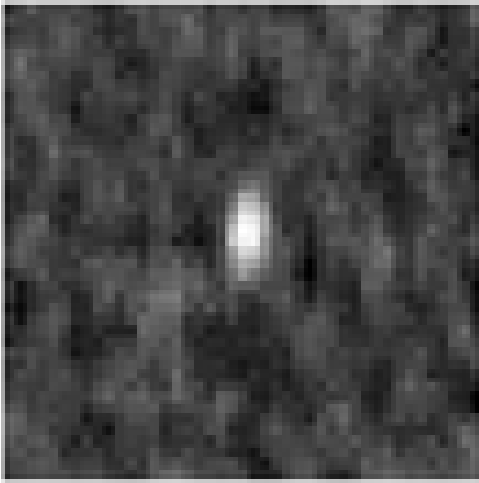}
\includegraphics[scale=0.5,ext=ps]{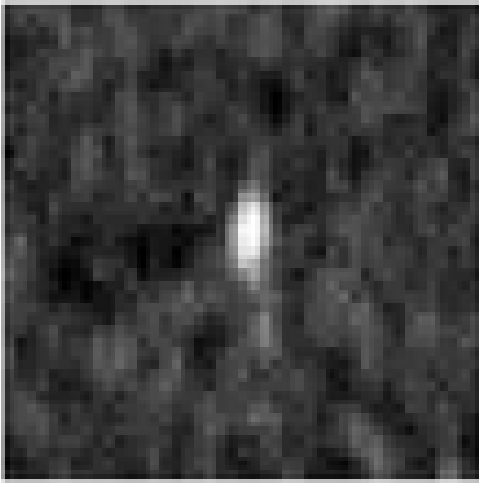}
\includegraphics[scale=0.5,ext=ps]{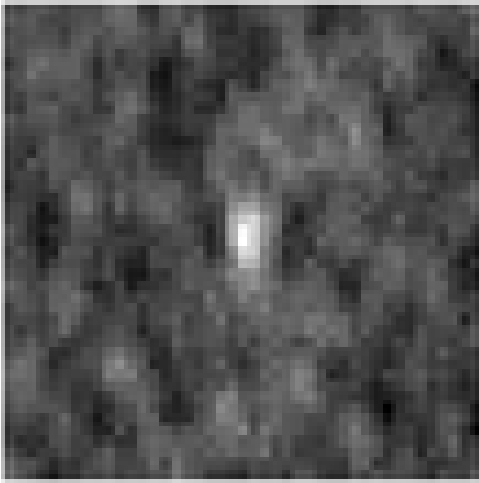}
\includegraphics[scale=0.5,ext=ps]{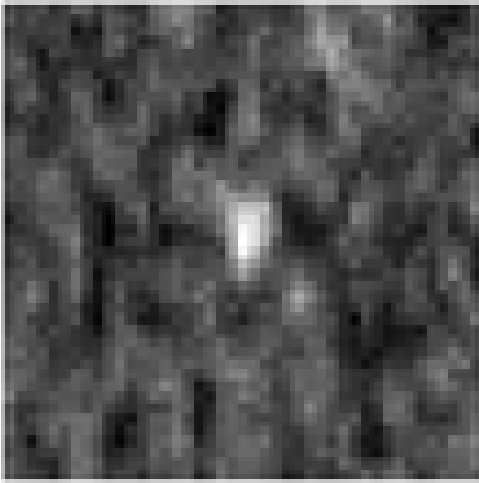}
\includegraphics[scale=0.5,ext=ps]{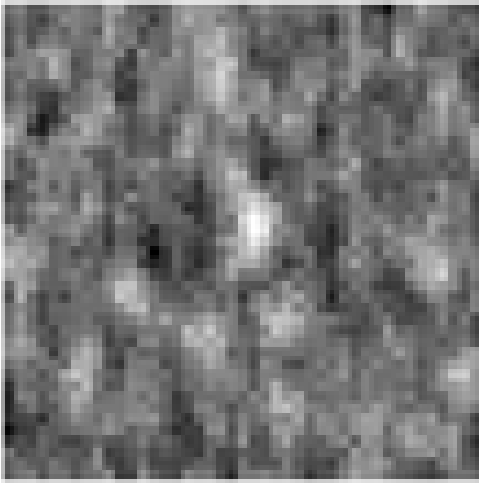}
\caption{Images from the {\it quiet-source} stacks. Upper left bin
  corresponds to faintest bin; $2< \log \i < 2.1$.  Bins
  increase $\log \i=0.1$ from left to right and top to bottom.}
\label{bins}
\end{figure}

\begin{figure}
\centering
\includegraphics[scale=0.5,ext=ps]{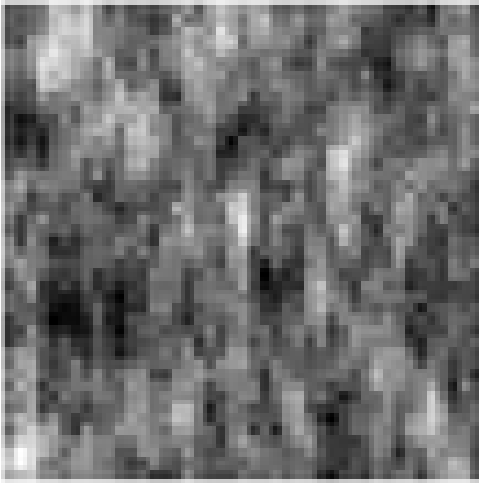}
\includegraphics[scale=0.5,ext=ps]{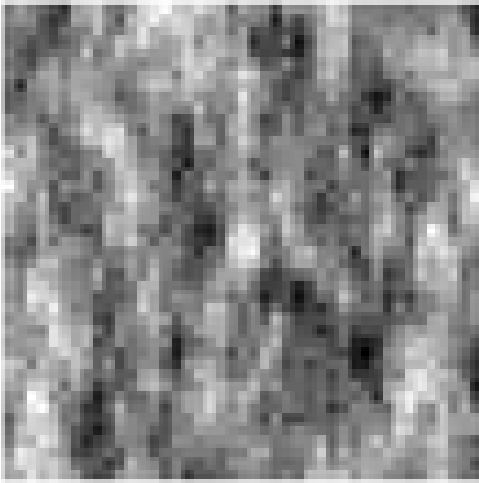}
\includegraphics[scale=0.5,ext=ps]{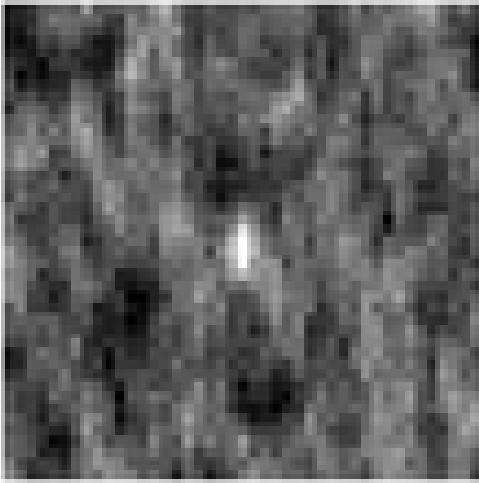}
\includegraphics[scale=0.5,ext=ps]{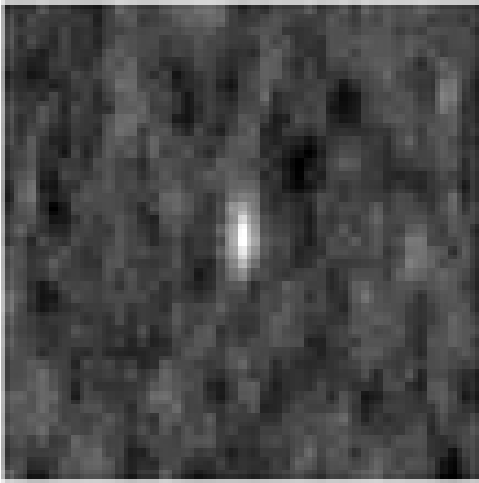}
\includegraphics[scale=0.5,ext=ps]{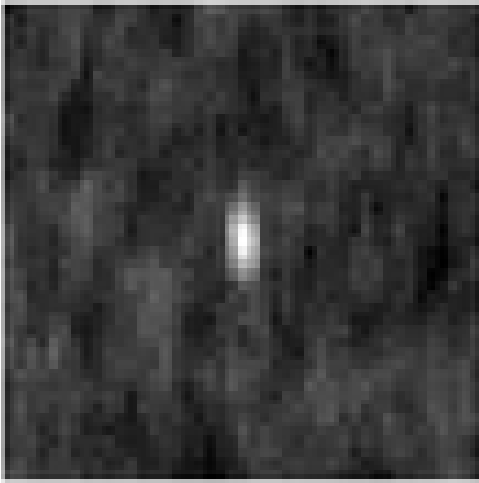}
\includegraphics[scale=0.5,ext=ps]{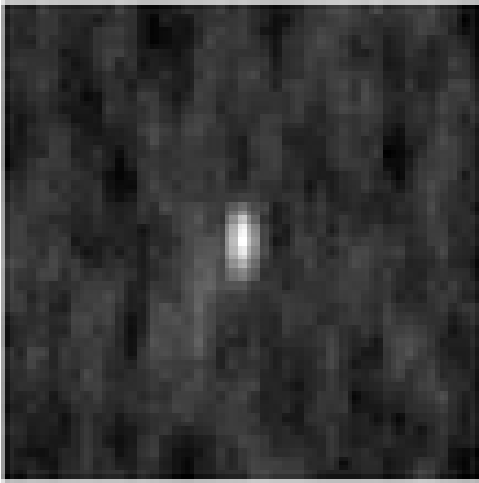}
\includegraphics[scale=0.5,ext=ps]{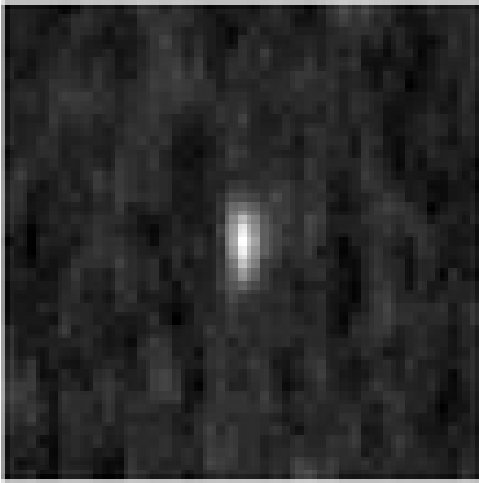}
\includegraphics[scale=0.5,ext=ps]{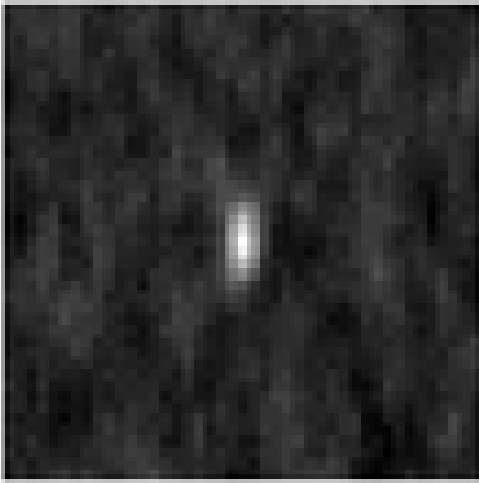}
\includegraphics[scale=0.5,ext=ps]{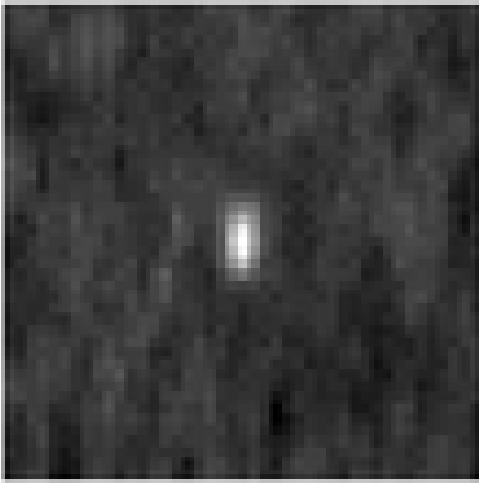}
\includegraphics[scale=0.5,ext=ps]{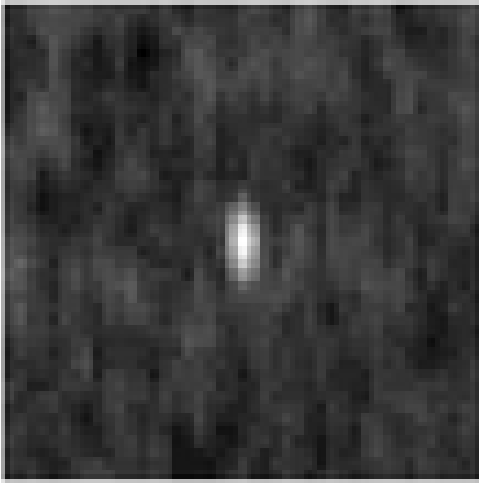}
\includegraphics[scale=0.5,ext=ps]{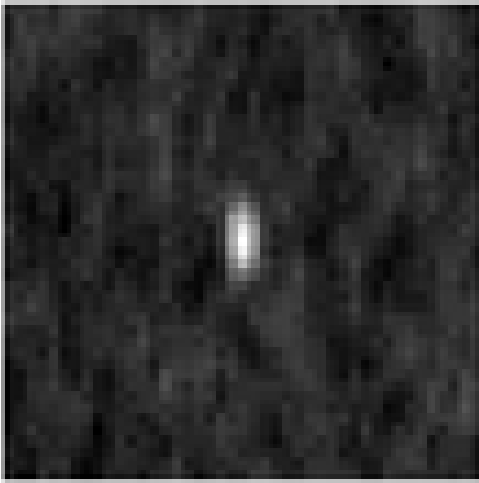}
\includegraphics[scale=0.5,ext=ps]{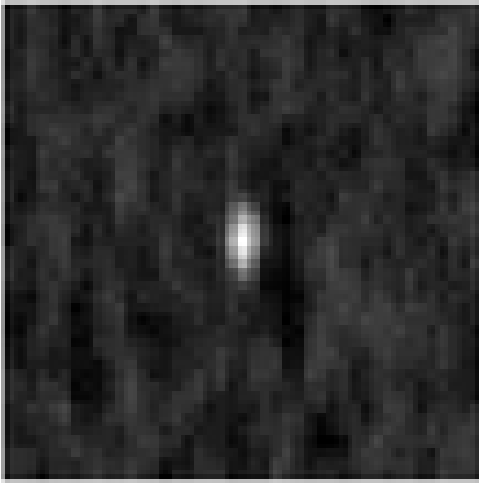}
\includegraphics[scale=0.5,ext=ps]{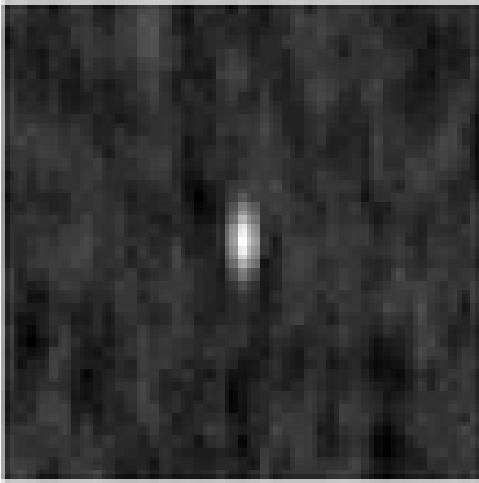}
\includegraphics[scale=0.5,ext=ps]{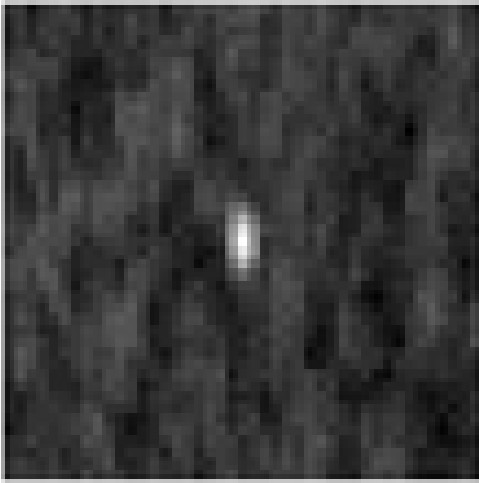}
\caption{Images from the simulated {\it quiet-source} stacks. Upper left bin
  corresponds to faintest bin; $2< \log \i < 2.1$.  Bins
  increase $\log \i=0.1$ from left to right and top to bottom.}
\label{simbins}
\end{figure}

\begin{table*}
\centering
\caption{Flux measurements of stacked images in the CDFS field}
\label{tab:flux}
\begin{tabular}{@{}rrrrrrr@{}}
\hline
&\multicolumn{3}{c}{{\it quiet-source} stack}&
\multicolumn{3}{c}{{\it all-source} stack}\\
$\i$& 
Number&
$\r$&
$\sigma$ (20cm)&
Number&
$\r$&
$\sigma$ (20cm)\\
($\mu$Jy)&&($\mu$Jy)&($\mu$Jy)&&($\mu$Jy)&($\mu$Jy)\\
\hline
                      115.2 & 200 & 5.91 & 2.57 & 205 & 6.37 & 2.55\\
                      145.1 & 568 & 4.88 & 1.60 & 581 & 5.21 & 1.61\\
                      180.8 & 1278 & 5.43 & 1.01 & 1317 & 6.90 & 1.01\\
                      225.3 & 1972 & 8.71 & 0.91 & 2039 & 10.03 & 0.90\\
                      281.0 & 2198 & 11.73 & 0.87 & 2295 & 13.60 & 0.89\\
                      351.5 & 1619 & 14.36 & 0.98 & 1685 & 15.77 & 0.98\\
                      442.2 & 1193 & 18.18 & 1.21 & 1267 & 21.42 & 1.17\\
                      556.6 & 791 & 21.45 & 1.62 & 841 & 25.59 & 1.57\\
                      699.4 & 479 & 25.64 & 1.90 & 527 & 30.65 & 1.97\\
                      870.3 & 296 & 33.70 & 2.57 & 325 & 37.48 & 2.47\\
                      1105.3 & 194 & 29.92 & 2.80 & 241 & 39.30 & 2.55\\
                      1383.7 & 120 & 38.01 & 3.82 & 151 & 54.13 & 3.44\\
                      1722.9 & 73 & 50.37 & 5.67 & 97 & 65.46 & 4.95\\
                      2250.8 & 30 & 56.59 & 7.30 & 54 & 91.15 & 5.31\\
                      2858.9 & 20 & 46.54 & 10.05 & 51 & 128.19 & 6.71\\
\hline
\end{tabular}
\end{table*}

\begin{figure}
\centering
\label{radir}
\includegraphics[scale=0.35,angle=270]{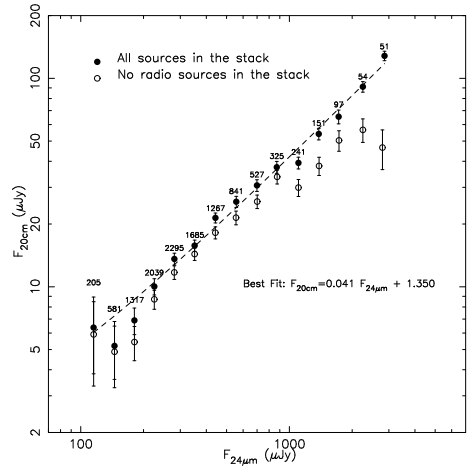}
\caption{The correlation between median radio flux, $\r$, and
infrared flux, $\i$, for the stacked CDFS sources.  Flux
measurements are included for stacks with (filled circles) and without
(open circles) the inclusion of detected ($>100\mu$Jy) radio flux at
the SWIRE sources positions. The numbers denote the numbers of SWIRE
source positions included in each stack).  The dashed line is the
best-fit line to the data obtained from the '{\it all-source}' stack.}
\end{figure}

\begin{figure}
\centering
\label{simradir}
\includegraphics[scale=0.35,angle=270]{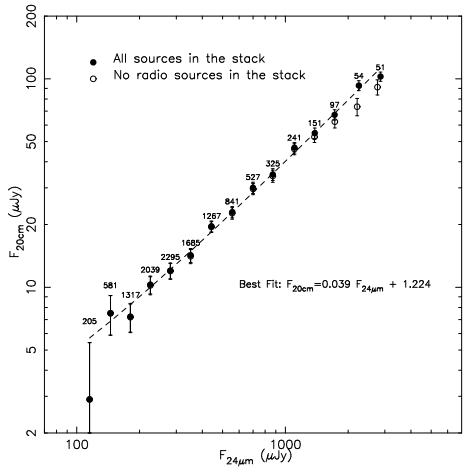}
\caption{The correlation between median radio flux, $\r$, and
infrared flux, $\i$, for the simulated data with the scaling
0.04.}
\end{figure}


\section {Stacking Analysis}

\subsection{Method}

We present here in detail the analysis of the radio and infrared data obtained in the CDFS and ELAIS field. Both fields were analysed separately.  A more detailed analysis was conducted on the CDFS field because of the more extensive data set (deeper SWIRE data, optical CCD data), but the ELAIS field provided a useful independent check of the results.  

We first removed galactic stars from the SWIRE catalogue in both fields.  For the CDFS we used the optical-infrared colour ($r-i$/$r-3.6\mu$m) selection method employed by Rowan-Robinson et al.\ (2004).  Optical magnitudes for all SWIRE sources were obtained from CTIO CCD observations (Siana et al.\ in preparation).  At bright optical magnitudes ($r<17$), image saturation renders the optical magnitudes unreliable and so the UKST R-band sky survey images of bright SWIRE ($\i>1000\mu$Jy) sources were inspected visually using the SuperCOSMOS sky server to remove any further stellar contamination.  For the ELAIS field we did not have access to reliable broadband optical magnitudes and so relied on the stellar classification provided by the SuperCOSMOS catalogue (confirmed by visual inspection for the brighter images).  Reliable star/galaxy classification is only possible at $r<20$mag from these measurements, and so it is possible that some residual stellar classification remains amongst the fainter optical counterparts.

Using {\sc AIPS++} we then extracted $43\times 43$ arcsec$^2$ sub-images from the 20 cm images, centred on each remaining SWIRE extragalactic source above the $5\sigma$ 24$\mu$m  flux limit in the CDFS ($\i > 100 \mu$Jy) and ELAIS ($\i > 400\mu$Jy) fields contained within the 48-arcmin radius region.  

The 20cm images were then sorted into logarithmic (0.1 dex) 24$\mu$m flux bins. We computed a stacked radio image for each bin by forming the median of corresponding pixels in all radio images in that bin. For each bin, two stacks were created: one which included all sources (the {\it all-source} stack) and one which excluded those images whose central pixels had $\r >100\mu$Jy (the {\it quiet-source} stack).  The radio flux for each bin was then estimated from the measured peak flux in the stacked image.  We also measured fluxes by fitting a point source to the image. This yielded identical results, indicating that
the stacked images were not resolved.  We also confirmed that stacking similar number of random positions within CDFS field produced no significant detections.  Images of the {\it quiet-source} stacks are shown in Fig~\ref{bins}.
  
To estimate the accuracy of the flux determinations, we created simulated radio images in AIPS++ with constant noise levels (rms=30$\mu$Jy) across the CDFS field. To this we added sources at the positions of all SWIRE sources with $\i>100\mu$Jy with a scaled, $\r=0.04 \i $, radio flux.  We repeated the analysis above, computing both {\it all-source} and {\it quiet-source} stacks.  The simulated
median images are shown in Fig~\ref{simbins}. As a further check on our analysis processes, we repeated the simulation by adding the artificial sources to simulated $u$--$v$ data, which we then processed in an identical way to the real $u$--$v$ data. This yielded an essentially identical result to the simulations that had been performed in the image plane.

\subsection{CDFS field}

The resultant fluxes for the stacked images in the CDFS are listed in Table~\ref{tab:flux}, together with the median infrared source flux $\i$ in each bin.  In many of the fainter $\i$ bins, we have been able
to stack over 1000 images.  However, we note that the rms in the resultant stacked image (estimated from the image by excluding the central quarter) has not decreased in line with Poisson statistics,
probably because of confusion from sub-$\mu$Jy radio sources. Assuming the infrared/radio correlation identified below, sources with $\i=10\mu$Jy will have $\r\sim 0.5 \mu$Jy, and extrapolating the $\i$ number counts (see Table~\ref{tab:flux}) predicts a source density of over $10^5$ sources deg$^{-2}$ at $\i=10\mu$Jy, approaching 10 sources per stacked image.  A confusion noise threshold of $\sim 0.5\mu$Jy when added in quadrature to an assumed Poissonian reduction in the noise afforded by the stacking procedure, i.e. $30\mu{\rm Jy}/\sqrt{n}$, is consistent with the observed rms noise measurements. 

The images from the {\it quiet-source} stacks in the CDFS field are shown in Fig~\ref{bins}, and the median values of $\i$ and $\r$ are plotted in Fig. 4. Corresponding images of simulated data are shown in Fig~\ref{simbins}, and their $\r$ {\it vs} $\i$ plots are shown in Fig. 5. The qualitative agreement between the corresponding real and simulated images is good but not excellent. The simulations do not reproduce the fine scale noise in the real data, which we believe is partially due to the residual effects of the bright source to the southwest. In support of this conclusion, we note that by moving the field centre closer to that source the level of the fine scale noise increases slightly in the stacked images.

Nevertheless, we note that the flux estimates from the simulated data (both the {\it all-source} and {\it quiet-source} stacks) are fitted by a linear relation

$$\r=0.044\i+4.74\mu {\rm Jy}\qquad $$

which is essentially identical to the input simulation. We conclude that the stacking introduces no systematic bias into our flux measurements.  

We note that the $\r$ {\it v} $\i$ correlation flattens off at the brightest flux bins for the {\it quiet-source} stack.  This reveals the systematic under-estimation of the radio flux in the {\it quiet-source}
stack caused by the exclusion of the increasingly large number of radio sources ($> 20$ per cent) at the $\r>100\mu$Jy level that are bona-fide counterparts to the SWIRE sources at high $\i$ fluxes.  Note
that this bias is also seen for the simulated {\it quiet-source} stacks,
although only significant for simulated stacks with $\i>2000\,\mu$Jy.
We obtain a fit to both the {\it all-source} stack (all bins) of the form:

$$\r=0.041(\pm0.002)\i+1.35(\pm 0.8)\mu{\rm Jy}$$

or an infrared-to-radio flux ratio, $q_{24}=\log(\i /\r) = 1.39\pm0.02$.  
The data points derived for {\it quiet-source} stack have a slightly lower normalisation throughout the infrared flux range.  This is because objects are detected above our radio limit in all $\i$ bins used in this analysis.  The fraction with radio detections rises from 2 per cent for the lowest $\i$ values to 40 per cent at the highest (see Table 1).

\begin{table*}
\centering
\caption{Flux measurements of stacked images in the ELAIS field}
\label{tab:elais_stack}
\begin{tabular}{@{}rrrrrrr@{}}
\hline
&\multicolumn{3}{c}{{\it quiet-source} stack}&
\multicolumn{3}{c}{{\it all-source} stack}\\
$\i$& 
Number&
$\r$&
$\sigma$ (20cm)&
Number&
$\r$&
$\sigma$ (20cm)\\
($\mu$Jy)&&($\mu$Jy)&($\mu$Jy)&&($\mu$Jy)&($\mu$Jy)\\
\hline
                      466.7 & 232 & 19.75 & 2.27 & 241 & 21.51 & 2.23\\
                      568.3 & 607 & 27.10 & 1.47 & 638 & 28.59 & 1.45\\
                      697.5 & 450 & 31.33 & 1.71 & 487 & 35.07 & 1.69\\
                      887.7 & 276 & 35.02 & 2.16 & 320 & 43.86 & 2.08\\
                      1100.4 & 122 & 46.06 & 3.20 & 156 & 59.80 & 2.96\\
                      1416.5 & 70 & 52.71 & 4.05 & 98 & 64.61 & 3.43\\
                      1703.8 & 41 & 51.89 & 5.53 & 53 & 65.55 & 4.77\\
                      2147.9 & 27 & 63.69 & 6.35 & 40 & 76.22 & 5.40\\
                      2811.5 & 13 & 53.32 & 8.61 & 28 & 121.77 & 6.28\\
 
\hline
\end{tabular}
\end{table*}

\begin{figure}
\centering
\label{fig:elaisstack}
\includegraphics[scale=0.35,angle=270]{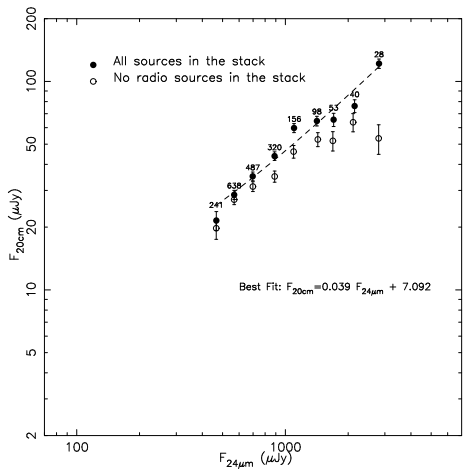}
\caption{The correlation between median radio flux, $\r$, and infrared flux, $\i$, for the stacked ELAIS sources.  Flux measurements are included for stacks with (filled circles) and without (open circles) the inclusion of detected ($>100\mu$Jy) radio flux at the SWIRE sources positions. The numbers denote the numbers of SWIRE source positions included in each stack).  The dashed line is the best-fit line to the data obtained from the '{\it all-source}' stack.}
\end{figure}

\subsection{ELAIS field}
Following the same procedures we obtained a value for the infrared-radio correlation using the ATCA and SWIRE observations of the ELAIS region. The ELAIS region does not extend to as deep a 24$\mu$m flux limit as the CDFS SWIRE field (see Table \ref {tab:elais_stack}).  A fit to the {\it all-source} stack gives the relation:

$$\r=0.039(\pm0.004)\i+7.1(\pm 3.3)\mu{\rm Jy}$$

for the {\it all-source} stack.   The slope is consistent with the result obtained from the CDFS data, giving a $q_{24}=1.41\pm 0.04$.  The intercept is different but this is largely dominated by fitting to the data points at the lowest flux levels where the errors are high and the fit poorly constrained.   Forcing the correlation to go through the origin of the $\i$/ $\r$ relationship gives an consistent fit in both cases with $q_{24}=1.40\pm0.02$. The correlation is shown in Fig.\ 6 and the fluxes for the stacked images are given in Table~\ref{tab:elais_stack}.

\section{Discussion}

\begin{figure}
\centering 
\label{ratio}
\includegraphics[scale=0.4,ext=ps]{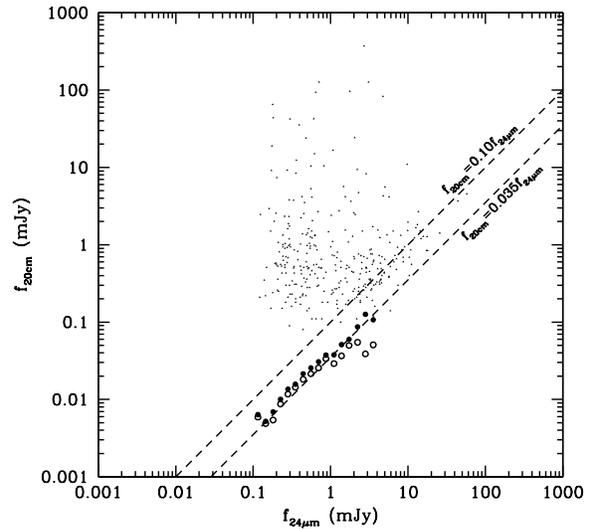}
\caption{A composite plot of the CDFS field shown $\i$ and $\r$ fluxes for individual sources (small triangles) and the {\it all-source} (filled circles) and radio-quiet (open circles) stacked images.}
\end{figure}

\begin{figure}
\centering 
\label{ind}
\includegraphics[scale=0.33]{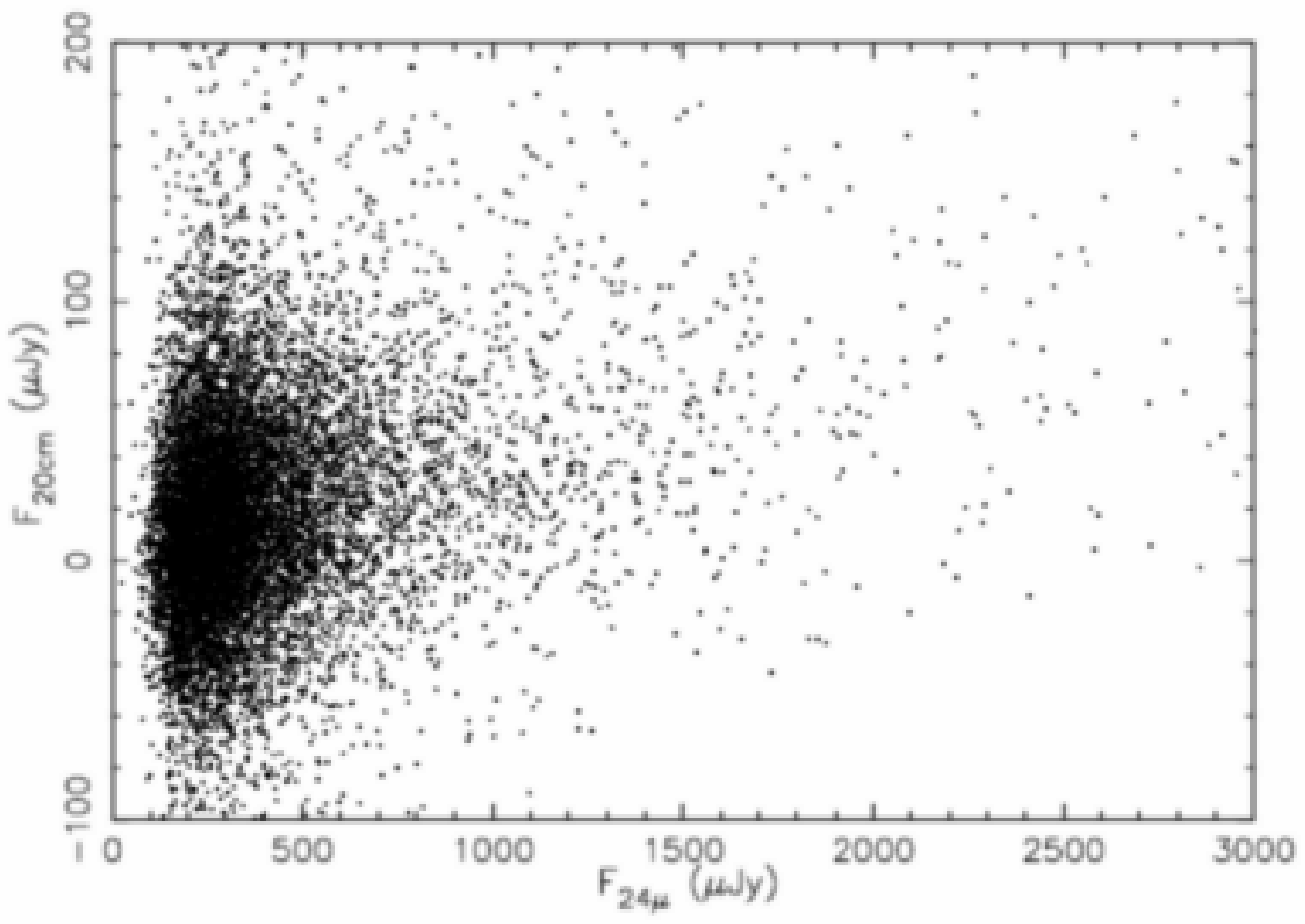}
\includegraphics[scale=0.33]{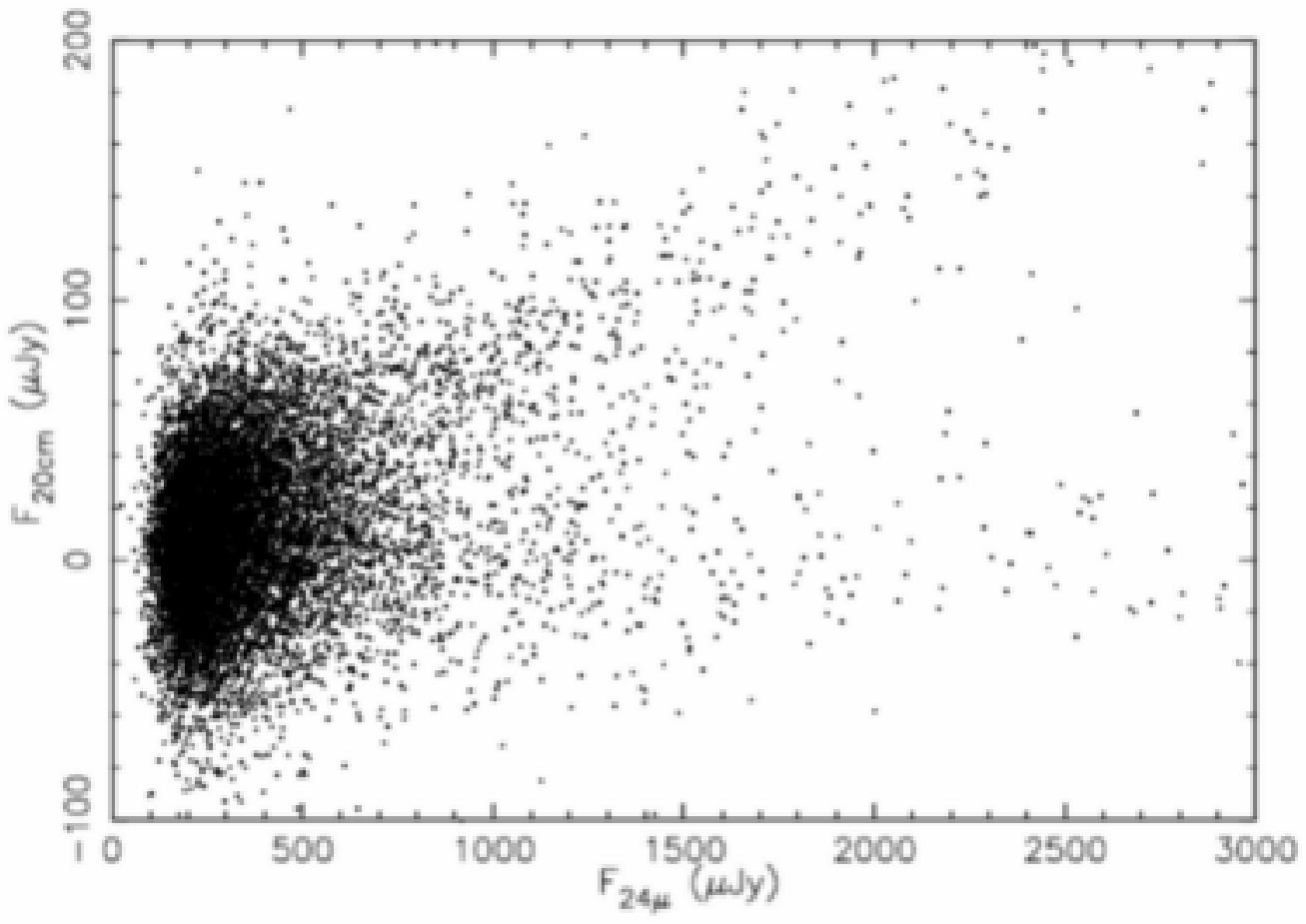}
\caption{Upper: Radio flux, $\r$, and infrared flux, $\i$, for the individual CDFS sources.  Fits using Lower: As for upper plot, with simulated sources: half with $\r=0$; half with $\r=0.04\i$, based on CDFS $\i$ distribution. See text for more details.}
\end{figure}

The most significant result to emerge from this analysis is that infrared-selected sources broadly follow the radio/far-infrared (FIR) correlation down to microJy flux densities, implying that these sources are primarily driven by star-formation rather than by AGN activity. This result rules out models such as that of Wall et al.\ (1986), who propose that the low-flux radio sources are low-luminosity radio galaxies.  Moreover, it confirms predictions that the far-infrared emitting starbursts should dominate the radio population at $\r < 50\mu$Jy (Gruppioni et al.\ 2003) based on extrapolations of models which fit the mid-infrared/radio correlation at brighter fluxes.

The results presented here show that, in broad terms, this same correlation holds for sources down to microJy levels, thus extending the range over which the correlation is known to hold by some two orders of magnitude in flux density. However, the data do not immediately tell us whether these galaxies are high-luminosity high-redshift star-forming galaxies or low-luminosity low-redshift star-forming galaxies.  

Most previous studies of the radio-FIR correlation have measured $q_{\rm FIR} = \log(S_{\rm FIR} /S_{\rm 20cm})$, where $S_{\rm FIR}$ is an integrated far infrared flux typically obtained from {\it IRAS} fluxes using a weighting algorithm e.g. $S_{\rm FIR} {\rm (Wm^{-2})} = 1.26 \times 10^4 ({2.58S_{60\mu{\rm m}}+S_{100\mu{\rm m}}}){\rm Jy}$.  Such studies, whether on radio-selected (Condon et al.\ 1991), infrared-selected (Condon \& Broderick 1991), or distance-limited (Helou et al.\ 1985) samples, all broadly give the same value of $q_{FIR}=2.3\pm0.2$.

Here we measure $q_{24} = \log(\i /\r)$, rather than $q_{FIR}$, so to compare our results with previous authors requires a conversion factor between $q_{24}$ and $q_{FIR}$, which will depend sensitively on the SED of the parent population and the MIPS filter response, and is beyond the scope of this paper. However, $q_{24}$ has been measured by Appleton et al.\ (2005) who obtain a (non-k-corrected) value of $q_{24}=0.84$, which differs significantly from our measured value of $q_{24}= 1.39$.  

Although our value of $q$ is derived from stacked data, whilst the Appleton value is derived from detected sources, this disrepancy is nevertheless a surprising result and merits further scrutiny.  First, we note that the simulations discussed in Section 3 have eliminated the possibility that it might be caused by an error in our imaging, stacking, or source-fitting algorithms or software.

Another possibility is that the discrepancy might be caused by a difference in k-correction between our analysis and that of Appleton et al.\ (2004). We have made no k-correction to this flux ratio because we have neither redshifts nor detailed spectral information on these objects in either the infrared or radio bands. Appleton et al.\ (2004) explicitly demonstrate  that individual k-corrections have little systematic effect on the mean trend of $q_{24}$ as a function of redshift at $z<1$. Based on spectral energy distribution modelling in the infrared, and an assumed synchrotron spectral index of $-0.7$ in the radio, Appleton et al.\ found that this increased the uncorrected value of $q_{24}=0.84$ by only 0.1--0.2, depending on the SED template used.   

However, Fadda et al.\ (2006) demonstrate that the k-correction arising from the PAH emission features entering the 24$\mu$m band at higher redshifts ($1.1<z<1.8$ or $2.2<z<2.5$) could be much greater.  Fadda et al.\ (2006) found that emission in the 24$\mu$m band could be enhanced by as much as a factor 2 ($\delta q=0.3$) for an M82 starburst galaxy template.  Thus it is possible that k-corection provides a partial explanation for the differences seen with Appleton et al.\ (2004), if the bulk of the population used in the stacking analysis are high redshift starbursts.  For k-correction to fully explain the difference our sources would need to be dominated by high redshift ($z>1$) starburst galaxies with PAH emission features approximately double that seen in some of the most intense star-forming galaxies locally.

A more mundane source for the discrepancy may arise from the different sample selection effects. We note that $q_{24}$ derived by Appleton et al.\ (2004) is based on radio detections of infrared sources.  Our values derived here are for all infrared sources, irrespective of radio flux.  Thus a lower coefficient could be explained by a population of sources with low radio-infrared flux ratios which have previously not been included in the study of radio-selected samples because they fall below the radio flux limit for the sample.

Indeed, we note that Gruppioni et al.\ (2003) found that the mid-infrared (15$\mu$m)/radio (20 cm) luminosity correlation for mid-infrared sources with radio detections was a factor of two higher than the correlation for all sources with upper limits for radio luminosities included in the analysis.  However, if this were to be the case we would still have to explain the difference between these lower $q_{24}$ (passband issues notwithstanding) and those FIR/radio correltions derived from {\it IRAS} sources with complete radio detections (Yun, Reddy \& Gruppioni 2003). 

\begin{figure*}
\centering 
\label{gal}
\includegraphics[scale=0.99]{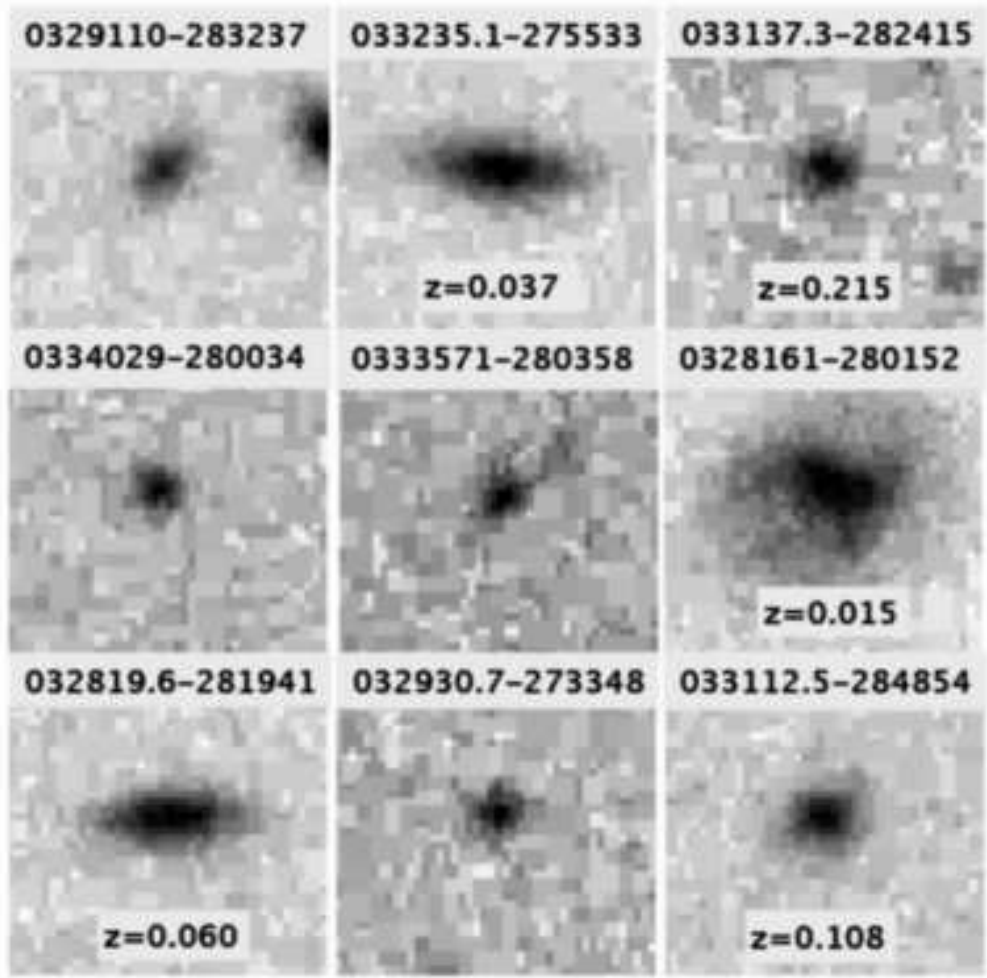}
\caption{$R$-band UK Schmidt telescope images of nine galaxies with anomalously low radio-infrared flux ratios.  Each image is 30$\times$30 arcsec. For top-left to bottom-right, the galaxies have $\i$ fluxes of 13.5, 6.99, 6.06, 5.97, 5.75, 5.29, 3.98, 3.97 and 3.86mJy, and all are below the 3$\sigma$ radio flux limit ($\r<100\mu$Jy). As indicated, four galaxies have spectroscopic redshifts from the 2dF Galaxy Redshift Survey (Colless et al.\ 2001).}
\end{figure*}

\begin{figure}
\centering 
\label{mips}
\includegraphics[scale=0.25]{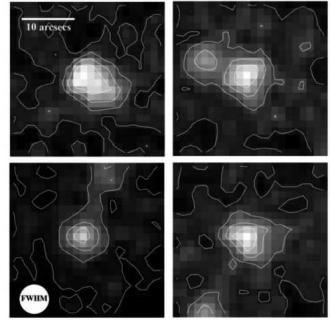}
\caption{MIPS $24\mu$m images for four CDFS source with $0.8< \i < 0.83$mJy and no detected 20cm emission.  Note the complex/extended morphology in the majority of cases, broadly consistent with the population at this flux level and $q_{24}$ value.}
 \end{figure}

To investigate this further, we plot in Fig. 7 a composite of the radio/infrared fluxes derived for both the individual radio-detected sources in the CDFS (Norris et al.\ 2006) and the stacked images derived in this paper.  The sample of detected sources includes both AGN and star forming galaxies, resulting in a significant tail in Fig. 7 extending to high radio-FIR ratios, and precluding a measurement of $q_{24}$. However there is a significant absence of sources below $\r=0.1\i$, (the radio-FIR relation found by Appleton et al.), which appears to mark a lower bound to the distribution. In particular, there is a significant absence of sources along the line extrapolated from our stacked data. Indeed, at the higher ($\i>10$mJy) fluxes nearly all non-stellar 24$\mu$m sources are detected at 20cm in the CDFS field ($\r > 0.4$mJy).  

As a further consistency check of our result we derived the infrared-radio correlation based on radio and infrared fluxes measured at the positions of individual sources used in our stacking sample. Most of the radio sources will lie below the formal 5$\sigma$ detection limit and consequently their fluxes will be subject to large errors.  Nevertheless the ensemble properties of the distribution will be robust.  In each infrared bin used above we computed the median value of the fluxes in each bin and made the fit to those points. For the CDFS and ELAIS the resulting fits were respectively:

$$\r=0.037(\pm0.001)\i+2.6(\pm 0.8)\mu{\rm Jy}\quad {\rm Median}$$
$$\r=0.041(\pm0.002)\i+6.6(\pm 3.5)\mu{\rm Jy}\quad {\rm Median}$$

The median estimator to the individual points in both fields gave a consistent slope to that obtained from the stacked images.  

We also modelled the effects of a population with a bimodal radio-infrared flux ratio; half of the population having $\r=0$ and half having $\r=0.08\i$ (based on the CDFS MIPS flux distributions). Although the stacked images of this simulated population give similar results to that obtained with the real data, the distribution of the individual simulated population flux values is visibly bimodal in a scatter plot of the individual source fluxes (Fig.~8 upper panel) and clearly different from a similar plot using the real data (Fig.~8 lower panel)

Nevertheless, we are faced with the conundrum that, whatever the distribution, the mean observed $q_{24}$ value is greater than that of any modelled SED.  Some of the optically-brightest CDFS sources with the most extreme $q_{24}$ values are relatively low redshift galaxies (Fig. 9).    However, most of these radio-weak sources have no direct optical counterpart at photographic plate limits ($R \sim 21\,$mag).   The majority of the objects at high $\i$ flux end of the correlation $\i \sim 0.83$mJy and show extended or complex morphologies in their $24\mu$m images (Fig. 10).  This would be consistent with a population dominated by actively star-forming galaxies at moderate redshifts.  To date, optical spectroscopy has been obtained for $\approx$ 130 SWIRE sources in the CDFS (Norris et al.\ 2007).  The median redshift is $z\sim0.4$ with a tail to higher redshift $z\sim 2$, exclusively composed of broad emission line active galactic nuclei at $z>1.2$.  However, this redshift distribution is strongly influenced by optical selection effects -- the vast majority of the objects with measured redshifts having relatively bright optical magnitudes ($R<22.5$\, mag).  

\section{Conclusions}

We find that the population of objects at $\mu$Jy flux densities obey a radio-24$\mu$m correlation which indicates that they are powered by star formation rather than by AGN, but their mean value of $q_{24}$ is inconsistent with previous results. We therefore propose that at microJy sensitivity, radio surveys are sampling a different population of objects, or objects with an intrinsically different value of $q_{24}$, from objects seen in the local Universe. Nevertheless, we acknowledge that the change obtained here for the mean value of $q_{24}$ occurs when we move from the use of discrete radio ($>5\sigma$) detections to fainter fluxes. Despite extensive testing and simulations we cannot rule out some unsuspected selection effect which may also contribute to this result. A partial explanation of the discrepancy may arise from the k-correction, but only if the population is dominated by extreme starburst galaxies at $1<z<1.8$ or $2.2<z<2.5$. Another explanation for at least part of the discrepancy may arise from the radio-selection biases inherent in some previous surveys.  Data from high-sensitivity surveys such as ATLAS will help to resolve such issues.

\section*{Acknowledgments}

Radio data was obtained by the Australia Telescope Compact Array,
operated by CSIRO Australia Telescope National Facility. IRS acknowledges support from the Royal Society.

\label{lastpage}

\end{document}